\documentclass{ws-p8-50x6-00}
  \begin{document}

  \title{On the Study of the Substructure of N-body Systems}

  \author{K.M.Bekarian, A.A.Melkonian}

  \address{Yerevan Physics Institute and Garni Space Astronomy Institute,
  Armenia}

  \maketitle

  \abstracts{
  Statistical properties
  of nonlinear systems revealed in an unexpected way already in Fermi-Pasta-Ulam 
paper, are responsible for the
  appearance of substructures.       
  We represent a method within S-tree diagram formalism for investigation of the 
substructures of N-body
  gravitating systems in 
              a defined time interval. This approach enables one to 
              consider the link between substructures of the system and its 
relative instability and is efficient for the
  study of the substructures
  in clusters of galaxies.} 

              \section{Introduction} 
              The study of hierarchical distribution of galaxies is one 
important issues of observational cosmology
  since the properties of substructure of galaxy clusters have to carry 
information on the mechanisms of their
  formation and evolution. 

  Various methods are developed for the study of the  substructures in galaxy 
clusters, e.g. 
              two and higher-point correlation functions, Li statistics, 
              topological measures, wavelets, etc which are using various 
              descriptors and assumptions 
              \cite{Peeb,Biv,Coma,Coles}. 
              $S$-tree method \cite{GK} is using both the positional and 
              kinematical information about the system in self-consistent way, 
by means of the consideration of the
  geometrical properties of the phase space of $N$-body system
  and the definition of the criterion of boundness between particles. 
  That method enabled to reveal the substructures in various galaxy clusters,   
i.e. Perseus, Virgo, Coma, Abell ENACS clusters, etc 
                         \cite{MN,GM,NewAstr}. 
              The existing versions of the 
              $S$-tree \cite{GK,BekMelk} are dealing with 
  the substructures at the fixed time. 

              In this paper we will study the behavior of substructure within  
              certain time interval i.e. to try to trace the future and past 
evolution of the subsystems within that
  interval. 

  The approach is based on the 
              generalized 
              $S$-scheme, and therefore, first, we will briefly review the main 
steps of the approach. 
    
              \section{N-body bounded systems: $S$-diagram method}

    The key idea of the method, as already mentioned above, is  the introduction 
of the 
    concept of the degree of boundness $\rho$. 

    Consider a set of $N$ points 

    $$ 
    X= \{x_1, \dots, x_N \}, 
    $$ 
    the function $P$ 
    $$ 
    P: X \times X \rightarrow R_{+} \quad \rm{and} \quad \rho \in R_{+}. 
    $$ 
  The basic definitions are as follows. 

    {\bf Definition 1} 

    We say that $\forall x \in X$ and $\forall y \in X$ are $\rho$-{\it 
bounded}, 
    if $P(x,y) \geq \rho$. 

    {\bf Definition 2} 

    We say  that $U \subset X (U \neq \emptyset)$ is a $\rho$-{\it bounded 
    subgroup}, if: 
    \begin{enumerate} 
    \item  $\forall x \in U$ and $\forall y \in \bar U \Rightarrow P(x,y) < 
\rho$; 

    \item  $\forall x \in U$ and $\forall y \in U \quad \exists x=x_{i_1}, 
    x_{i_2}, \dots, 
    x_{i_k}=y$, 
    that $P(x_{i_l}, x_{i_{l+1}}) \geq \rho; \quad  \forall l=1, \dots,k-1$. 
    \end{enumerate} 
    {\bf Definition 3} 

    We say  that $U_1, \dots, U_d$ is the {\it distribution} of the set $X$ {\it 
via} 
    $\rho$-bounded groups, if: 
    \begin{enumerate} 
    \item  $\bigcup_{i=1}^{d}U_i=X;$ 
    \item $i \neq j \quad (i,j=1, \dots, d) \Rightarrow U_i \bigcap U_j= 
\emptyset;$ 
    \item $U_i (i=1,\dots,d)$ is a $\rho$-bounded group. 
    \end{enumerate} 
    The function $P$  Different physical quantities can be assigned to the 
function $P$ as 
  listed in \cite{GK}, such as the distances, the energy, the force, the 
    perturbations of 
    potential, momentum, and so on. For example, the 
    choice 
    of the mutual distance of the particles in order to define the subgrouping 
    is equivalent to the corresponding correlation functions 
    (spatial or angular, related with each other via the Limber equation), 
    which is, obviously, a rather incomplete characteristic for that aim. 
     It can be shown that, at least for astrophysical problems, among the most 
    informative ones is the Riemannian curvature of the configuration space 
    \cite{GK}, 
    which determines the behavior of close geodesics, as known from  basic 
    courses on classical mechanics \cite{Arn}. 

    So, by this, S-tree algorithm we obtain the substructure of the N-body 
system, 
  i.e. the degree of boundness of each subgroup of particles, for 
    any given function $P$ and $\forall \rho$. This found distribution will 
satisfy 
    Definitions 1, 2 and 3. The final result can be 
    represented either through 
    tables or by tree-diagrams, graphs (S-tree).     
      
              The next important aspect is that the method allows to use 
              simultaneously various boundness criteria between the particles of 
the system for fixed moment of time 
              \cite{BekMelk}. 
    
              Consider the $N$-body system: $X=\{x_1,x_2,\dots, x_N\}$ and $P_1, 
              \dots, P_H$ 
              functions. Consider also $D_1, 
              \dots, D_H$ 
              matrices, where $D_{\alpha}=(d_{ij}^{\alpha}), 
              \alpha=1, \dots, H$;\\ 
              $i,j=1,\dots, N; \quad d_{ii}^{\alpha}=0$. 
              If $i \neq j \quad d_{ij}^{\alpha}=P_{\alpha}(x_i,x_j)$. 
              Constructing matrix $D$ in the following way: 
              $$ 
              D=(\bar d_{ij}), \quad i,j=1, \dots, N, 
              $$ 
              where 
              $$ 
              \bar d_{ij}=(d_{ij}^1,d_{ij}^2,\dots, d_{ij}^H), \quad 
i,j=1,\dots, 
              N. 
              $$ 
              $D$ matrix contains the whole information about matrices 
$D_1,\dots, 
              D_H$. 
              For any function $P_{\alpha} \quad (\alpha=1, \dots, 
              H)$ 
              exists its own boundness vector \\ $\rho_{\alpha}= 
              (\rho_1^{\alpha},\dots,\rho_{Q_{\alpha}}^{\alpha}).$ 
              We construct for the $D$ matrix a vector of boundness 
              $$ 
              \bar \rho_D=(\bar \rho_1^D,\dots,\bar\rho_{Q_D}^D) 
              $$ 
              and for $k=1,\dots, Q_D;\quad 
              \bar\rho_k^D=(\rho_{k_1}^1,\dots,\rho_{k_H}^H)$, 
              where $k_{\alpha}=1,\dots,Q_{\alpha}$ and $\alpha=1,\dots,H$. 
              Vector $\bar \rho_D$ is constructed by all possible combinations 
of 
              components 
              of $\bar\rho_k^D$ vector. 
              For any function $P_{\alpha},\quad (\alpha=1,\dots, H)$ we 
introduce  
              "the degree of influence" $W_{\alpha}\in R_+$. 
              The next steps are realized for fixed value of $\bar\rho_k^D$. 
              We transit from the matrix $D$ to the matrix $D_u$ in a following 
way 
              $$ 
              D_u=(\bar u_{ij}); \quad i,j=1,\dots,N, 
              $$ 
              where $\bar u_{ij}=(u_{ij}^1,\dots,u_{ij}^H)$ and 
              $u_{ij}^{\alpha},\quad \alpha=1,\dots,H$ 
              is defined as 
              $$ 
              u_{ij}^{\alpha}=\left\{\begin{array}{c} 
              W_{\alpha}, \quad if \quad d_{ij}^{\alpha}\geq 
              \rho_{k_{\alpha}}^{\alpha}\\ 
              0, \quad if \quad d_{ij}^{\alpha} < \rho_{k_{\alpha}}^{\alpha}\\ 
              \end{array}\right. 
              $$ 
              The following $D_v$ matrix we construct as 
              $$ 
              D_v=(v_{ij}),\quad i,j=1,\dots,N, \quad 
              v_{ij}=\sum_{\alpha=1}^Hu_{ij}^{\alpha}. 
              $$ 
              Defining $\bar \mu=(\mu_1,\dots,\mu_S)$, the vector of boundness 
              for the $D_v$ 
              matrix and, using the $S$-method for $D_v$ and for any given 
component 
              of 
              $\bar \mu$ vector, we obtain the distribution of the initial 
system on the 
              bounded 
              subgroups. 
              Note finally, that according to the $S$-generalized scheme with 
fixed 
              values of 
              $\bar\rho_k^D$ and $\mu_l$, we obtain the distribution of system 
on 
              $\tilde \rho$-bounded subgroups, where $l=1,\dots,S;\quad 
              \tilde\rho=(\bar\rho_k^D;\mu_l)$. 

              Consider a dynamical time interval or including $z$ dynamical 
              interval 
              $[t_a,t_b]$ of an initial system \cite{Dis}. Divide this interval 
on $H-1$ equal 
              parts. 
              For any $t_{\alpha}; \alpha=1,\dots,H$ it is possible to 
              find all sets 
              of $(L_i(t_{\alpha}),V_i(t_{\alpha}))$, and where 
$i=1,\dots,N;\quad 
              L_i, V_i$ - 
              coordinates and velocities of $i$--th particle. 
              Consider the  $P$ function at different moments of time. 
              Represent the function $P_1,\dots,P_H$ as 
              $P_1=P(t_1)=P(t_{\alpha}), \quad 
P_2=P(t_2),\dots,P_H=P(t_H)=P(t_b).$ 
              
              As in the previous section, we will obtain $D$ and $D_u$ matrices. 

              For $W_{\alpha}=1, \alpha=1,\dots H, W_{\alpha}=1$. 
              For the matrix $D_v$ and $\mu \in R_+$ using the $S$-tree method 
we obtain the 
              splitting of the system into 
              on $\mu$-bounded subgroups within the about 
              time interval $[t_a,t_b]$. 
    
              As a prefered floor of distribution $\mu_*$ we chose the     
larger components of the vector $\bar\mu$,
  because $\bar\mu$-connection is a 
              quantitative 
              estimation of the sequence of numbers, where 
              $$ 
              u_{ij}^{\alpha}=1, \quad i,j=1,\dots,N, \quad \alpha=1, \dots,H 
              $$ 
              After splitting the system into bounded groups by the 
              suggested model, 
              $\Phi/N$ value may present a define physical 
              interest, where $\Phi$ -- is numebr of those 
              particles, which are not single in their groups 
              \cite{Dis}. 
              
          Consider two different systems $X_1, X_2$ with $N_1$ and $N_2$ 
              particles. 
              For both systems a distribution is analyzed, $[t_a^1,t_b^1]$ and 
              $[t_a^2,t_b^2]$ intervals have to be considered in a self-
consistent way. 
              The difference of $\Phi_1/N_1$ and $\Phi_2/N_2$ can be 
              considered as a 
              criterion of relative instability of these systems. 
              Note, that the presented approach allows to link directly the 
              substructure properties and the relative instability of the N-body 
systems. 
              
  In order to apply this scheme for physical systems the correspondence between 
the chosen time intervals
  $[t_a^1,t_b^1]$ and 
              $[t_a^2,t_b^2]$ have to be defined taking into account 
              the formulation of the criterion of relative instability for 
systems $X_1$ 
              and $X_2$.

              \section{Conclusion} 

  The study of the substructure of nonlinear N-body systems is an interesting 
  theoretical problem, as well has a direct link with the astrophysical systems. 
  Particularly, the S-tree approach had proved its practical efficiency in the 
study of the substructure 
  of clusters of galaxies. Particularly, the subgroups of the Coma cluster 
detected 
  by S-tree \cite{NewAstr} have been later studied observationally and certain 
predicted 
  trends in the properties of galaxies of one of subgroups (subgroup 2) have 
found their 
  apparent confirmation \cite{Mario}. 
  The further development of the method therefore, seems of particular interest. 
              
  Here we proposed a generalization of the approach enabling the investigation 
of evolution of the
  substructuring properties of a nonlinear system within a defined time
              interval. Application of this scheme to the clusters of galaxies 
can 
  lead to a valuable insight on the evolution and maybe on the origin of the 
substructures. 
                          
We are grateful to Prof.V.G.Gurzadyan for discussions and advice.

  \end{document}